\documentclass[aps,prb,preprint,showpacs]{revtex4}
\usepackage{graphicx}
\usepackage{dcolumn}
\usepackage{bm}

\begin{document}

\title{Theory of electronic transport in random alloys with
  short-range order: Korringa-Kohn-Rostoker non-local coherent
  potential approximation}

\author{P. R. Tulip}
\affiliation{Department of Physics, University of Warwick, Coventry,
  CV4 7AL, United Kingdom}
\author{J. B. Staunton}
\affiliation{Department of Physics, University of Warwick, Coventry,
  CV4 7AL, United Kingdom}
\author{S. Lowitzer}
\affiliation{Department Chemie und Biochemie, Physikalische Chemie,
  Universit\"{a}t M\"{u}nchen, Butenandtstrasse 5-13, D-81377
  M\"{u}nchen, Germany}
\author{D. K\"{o}dderitzsch}
\affiliation{Department Chemie und Biochemie, Physikalische Chemie,
  Universit\"{a}t M\"{u}nchen, Butenandtstrasse 5-13, D-81377
  M\"{u}nchen, Germany}
\author{H. Ebert}
\affiliation{Department Chemie und Biochemie, Physikalische Chemie,
  Universit\"{a}t M\"{u}nchen, Butenandtstrasse 5-13, D-81377
  M\"{u}nchen, Germany}

\begin{abstract}
We present an ab-initio formalism for the calculation of transport properties in
compositionally disordered systems within the framework of the
Korringa-Kohn-Rostoker non-local coherent potential approximation. Our
formalism is based upon the single-particle Kubo-Greenwood linear
response and provides a natural means of incorporating the effects of
short-range order upon the transport properties. We demonstrate the
efficacy of the formalism by examining the effects of short-range
order and clustering upon the transport properties of disordered $AgPd$ and $CuZn$ alloys.
 
\end{abstract}


\maketitle

\section{Introduction}

The Korringa-Kohn-Rostoker coherent potential approximation (KKR-CPA)
 \cite{cpa} represents an extremely successful one-electron theory capable of
describing the properties of many compositionally disordered alloy
systems. In particular, in combination with density functional theory
(DFT) \cite{dft1, dft2, johnson, johnson-totalE} and the Korringa-Kohn-Rostoker method
of band theory \cite{gyorffy, stocks, gyorffy2}, it provides a fully first-principles
description of such systems. A long history of successful applications
\cite{staunton, luders, faulkner, gyorffy3,hughes,MAEvsT} attests to the utility and
accuracy to which the method is capable. 

Of particular relevance to our
discussion here, the KKR-CPA has been used to calculate the transport properties
of such alloys. Historically, the early work concerning this topic
involved the use of a Boltzmann equation \cite{butler1, butler2,
  butler3}, and although these works demonstrated remarkable agreement
with experiment, they did suffer from two notable defects: namely, the requirement that well-defined energy bands exist 
within the alloy, and the
neglect of vertex, or ``scattering-in'', terms. The former defect
limits the application of such a theory to weak-scattering alloys
only, while the latter could be expected to lead to significant error
in systems such as those where appreciable s-p or s-d scattering
manifests itself.      

Velicky \cite{velicky} developed a CPA theory for transport using the
Kubo-Greenwood \cite{Kubo, Greenwood} formalism as his starting point, which was
applied to a two-level tight-binding Hamiltonian. Although capable of
yielding vertex corrections, this approach suffered from difficulties
when applied to realistic systems. For example, the necessity of
assuming that the wavefunctions are identical on all lattice sites,
irrespective of the occupying atomic species. The seminal work of Butler \cite{butler4}, resolved these issues, as
he developed a KKR-CPA theory based upon the Kubo-Greenwood linear
response formalism \cite{Kubo, Greenwood}. As a multiple-scattering
based approach, this did not require the existence of well-defined
energy bands, and further, Butler demonstrated how the vertex
corrections arose quite naturally within his formalism. Although the
formal developments of this work followed Velicky's quite closely, the
use of a realistic single-electron muffin-tin Hamiltonian allowed
connection with first-principles methods to be made. The method has
been applied with success to a range of alloy systems \cite{swihart,
  banhart} and it has also been successfully extended to the
relativistic regime \cite{banhart2}. 

Of course, all of these calculations suffer from the main drawback of
the CPA; namely that as a single-site mean-field theory, it is
incapable of incorporating the effects of local-environment
fluctuations in the alloy crystal potential. The CPA therefore
explicitly ignores the effects of such short-range order
(SRO) effects upon the physics of disordered alloys. As discussed by
Gonis \cite{gonis}, these statistical fluctuations can be
important. Whilst in general the presence of SRO is likely to diminish the
resistivity there are examples, the so-called ``Komplex'' K-state alloys \cite{K-state}, where the
onset of SRO is accompanied by an increase in the alloy residual
resistivity.

Recently there have been some successful attempts at calculating the effects of SRO upon
the electronic structure of disordered alloys and they provide the means to study transport properties.
Mookerjee and Prasad \cite{mookerjee,dasgupta,dasgupta2}, using a TB-LMTO method \cite{tb}
in conjunction with an augmented space formalism \cite{mookerjee2,kaplan} and real space recursion
method \cite{haydock} described SRO effects on alloy electronic densities of states and related quantities whilst Saha \emph{et al.} \cite{saha} have obtained
spectral functions within the same framework. Recently Tarafder \emph{et al.}~\cite{Tarafder}  have developed a formalism
for the optical conductivity and reflectivity from the same basis and used it to study copper-zinc alloys.
To date however it has not been possible to incorporate this technique fully within electronic
density functional theory. The recent work of Rowlands
\emph{et al} \cite{derwyn, derwyn2,derwyn dft}, along with Biava
\emph{et al.} \cite{biava, biava2} concern a development which is not restricted in this way. References~\onlinecite{derwyn, derwyn2}
formulate and illustrate a successful method for incorporating the effects of SRO
within the framework of KKR-CPA theory, whilst implementation for realistic
systems is described in references \onlinecite{derwyn2,biava}. The method can be readily combined with density functional theory to provide a first principles description of disordered alloys as demonstrated in  reference \onlinecite{derwyn dft, r-nlcpa}. This nonlocal CPA (NLCPA)
theory is based upon reciprocal space coarse-graining ideas introduced by Jarrell and Krishnamurthy
\cite{jarrell and krishnamurthy}
originating from the dynamical cluster approximation (DCA)
\cite{hettler, hettler2, maier}. The KKR-NLCPA \cite{derwyn, derwyn2} introduces an effective
(translationally invariant) disorder term
$\delta G$ which represents an effective propagator that accounts
for all nonlocal scattering correlations on the electronic propagation
due to disorder configurations and modifies the structure constants
accordingly. By coarse-graining reciprocal space, one naturally introduces real space periodically
repeating clusters. As such, the NLCPA maps an effective lattice
problem to that of an impurity cluster embedded in a
self-consistently determined effective medium, and thus yields a
cluster generalisation of the KKR-CPA that includes non-local
correlations up to the range of the cluster size. Unlike other cluster approaches, such as the molecular CPA
(MCPA) \cite{tsukada} it is fully translationally invariant, that
is, the effective medium has the site-to-site translational invariance
of the underlying lattice. It is also computationally tractable,
largely on account of the reciprocal space coarse-graining procedure
employed.

Thus far, the NLCPA has been employed to investigate the effects of
SRO upon the electronic structure of a range of realistic systems,
using both the muffin-tin Hamiltonian \cite{derwyn2, spectral}, and
the tight-binding approach \cite{gary}. Given its ability
to address such issues successfully and its proven incorporation into 
an electronic density functional theory~\cite{derwyn dft} (DFT) for disordered systems with SRO it makes sense then to extend it
to the calculation of transport properties. To this end, by invoking time dependent DFT ~\cite{TDDFT} within the adiabatic approximation, in this paper
we present a formalism for the determination of the residual
resistivity in the KKR-NLCPA, and explicitly demonstrate the efficacy
of the method through application to several realistic alloy
systems. Our theoretical formalism is a careful generalisation of that
of Butler \cite{butler4} and where appropriate we omit the steps in the
derivation which can be straightforwardly obtained from this paper.
 Our paper is structured as follows: The next section gives a short overview of the
conductivity tensor and then the transport coefficients available from the Kubo-Greenwood
formalism. This is followed by a section containing the salient points of the
KKR-NLCPA formalism including its use in implementing the density functional theory. We then develop
our theory for the conductivity of disordered systems with short-range order which
includes the treatment of `vertex corrections'. The implementation strategy is 
outlined before calculations for the effects of SRO on the resistivity of both
b.c.c. $CuZn$ and f.c.c. $AgPd$ are presented. 

\section{The Conductivity; Kubo-Greenwood linear response}

The Kubo-Greenwood \cite{Kubo, Greenwood} linear response formalism
states that, for disordered system, the symmetric part of the conductivity tensor
has coefficients $C$ which can be determined from the evaluation of
an expression of the form
\begin{equation}
C=\mbox{Tr}<O_{1}GO_{2}G>,
\end{equation}
where $G$ is a single-particle Green's function, which is dependent
upon the details of the effective one-electron potential. $O_{1}$ and $O_{2}$ are
operators, and the angled brackets denote a configuration average over the
distribution of the potentials. 

To determine the d.c. conductivity, we consider the following expression
~\cite{Kubo,Greenwood,butler4}
\begin{equation}
\label{greenwood equation}
\sigma_{\mu\nu}(E_{F})=\frac{\pi\hbar}{N\Omega}\bigg<
\sum_{mn} J_{mn;\mu}J_{nm;\nu}\delta(E_{F}-E_{m})\delta(E_{F}-E_{n})\bigg>,
\end{equation}
where $J_{mn;\mu}=\langle m|J_{\mu}|n\rangle$ denotes the matrix element of the
current operator in the $\mu$th spatial direction, which is given by
\begin{equation}
J_{\mu}=-i\hbar\frac{e}{m}\frac{\partial}{\partial r_{\mu}},
\end{equation}
with $|m\rangle$ and $|n\rangle$ denoting eigenfunctions of a particular configuration of the
disordered system. Here, $N$ is the number of atoms, and $\Omega$ is the
volume per atom. $E_F$ is the Fermi energy.

Within the KKR approach, the electronic structure of the alloy is
expressed in terms of the single-particle Green's function, rather
than in terms of eigenstates and eigenvalues of the Hamiltonian; we
can introduce the Green's function simply by using the identity \cite{henry}
\begin{equation}
-\pi\sum_{n}|n\rangle \langle n|\delta(E-E_{n})=\lim_{\eta \to 0}\mbox{Im}G(E+i\eta),
\end{equation}
while the awkward imaginary part of the Green's function may be
removed by writing
\begin{equation}
- 2i\pi\sum_{n}|n\rangle\langle n|\delta(E-E_{n})=\lim_{\eta \to 0}[G(E+i\eta)-G(E-i\eta)].
\end{equation}

Inserting this into Eq. (\ref{greenwood equation}), yields the following for the conductivity
\begin{eqnarray}
\sigma_{\mu\nu} & = & 1/4\lim_{\eta\to
  0}[\tilde{\sigma}_{\mu\nu}(E^{+},E^{+})-\tilde{\sigma}_{\mu\nu}(E^{-},E^{+})-\nonumber \\
& & \tilde{\sigma}_{\mu\nu}(E^{+},E^{-})+\tilde{\sigma}_{\mu\nu}(E^{-},E^{-})]
\end{eqnarray}
where we define the complex energies as
\begin{equation}
E^{+}=E_{F}+i\eta, \: E^{-}=E_{F}-i\eta, \: \eta \to 0
\end{equation}
and
\begin{equation}
\tilde{\sigma}_{\mu\nu}(z_{1},z_{2})=-\frac{\hbar}{\pi
  N\Omega}\mbox{Tr}<J_{\mu}G(z_{1})J_{\nu}G(z_{2})>,
\end{equation}
where $z_{1}$ and $z_{2}$ are each either $E^{+}$ or $E^{-}$.

For non-overlapping effective single electron potentials the Hamiltonian takes the form
\begin{equation}
H=-\frac{\hbar^{2}}{2m}\nabla^{2}+\sum_{i}v_{\gamma}(\mathbf{r}-\mathbf{R}_{i}),
\end{equation}
where the atomic positions $\mathbf{R}_{i}$ are fixed, and form a
regular lattice. The potentials $v_{\gamma}(\mathbf{r}_i)$ vary from site to
site ($\mathbf{r}_i = \mathbf{r}-\mathbf{R}_{i}$) and $\gamma$ is a configuration label. 

Within the multiple-scattering theory, the single-particle
Green's function for a given configuration $\gamma$ can be written as~\cite{faulkner and stocks}
\begin{widetext}
\begin{eqnarray}
\label{green function}
G_{\gamma}(E,\mathbf{r}_{i},\mathbf{r}^{\prime}_{j})=2m/\hbar^{2}\bigg(\sum_{\Lambda, \Lambda^{\prime}}Z^{i}_{\Lambda,\gamma}(E,\mathbf{r}_{i})\tau^{ij}_{\gamma; \Lambda,\Lambda^{\prime}}Z^{j, \dagger}_{\Lambda^{\prime}, \gamma}(E,\mathbf{r}^{\prime}_{j})-\sum_{\Lambda}Z^{i}_{\Lambda,\gamma}(\mathbf{r}^{<}_{i})J^{i, \dagger}_{\Lambda, \gamma}(E,\mathbf{r}^{>}_{i})\delta_{ij}\bigg),
\end{eqnarray}
\end{widetext}
where $\tau^{ij}_{\gamma,\Lambda,\Lambda^{\prime}}$ is the scattering path operator (SPO)
describing propagation between sites $i$ and $j$ in configuration $\gamma$,
$Z^{i}_{\Lambda,\gamma}(E,\mathbf{r}_{i})$ is the regular solution to the
Schr\"{o}dinger/Dirac equation in the cell surrounding the atom $i$, and
$J^{i}_{\Lambda, \gamma}(E,\mathbf{r}^{\prime}_{i})$ represents the irregular
solution within the same cell (note
that there should be no confusion with the current matrix elements
here). $\Lambda$ encapsulates the appropriate angular momentum quantum numbers.~\cite{butler4,banhart2}

In calculating the conductivity, the second term in the Green's
function expression (Eq.( \ref{green function})) is real and may
be omitted, when calculated for a real potential
at a real energy \cite{butler4}. Thus the conductivity may be written
\begin{widetext}
\begin{equation}
\label{tilde sigma}
\tilde{\sigma}_{\mu\nu}(z_{1},z_{2})=-\frac{4m^{2}}{\pi
  N\Omega\hbar^{3}}\sum_{i,j}\sum_{\Lambda_{1},\Lambda_{2},\Lambda_{3},\Lambda_{4}}<J^{i,\gamma}_{\Lambda_{1}
\Lambda_{2}, \mu}(z_{2},z_{1})\tau^{ij}_{\gamma;\Lambda_{2}\Lambda_{3}}(z_{1})J^{j,\gamma}_{\Lambda_{3}\Lambda_{4}, \nu}(z_{1},z_{2})\tau^{ji}_{\gamma; \Lambda_{4}\Lambda_{1}}(z_{2})>
\end{equation}
with 
\begin{equation}
J^{i,\gamma}_{\Lambda \Lambda^{\prime},\mu}(z,z^{\prime})=-\frac{ie\hbar}{m}\int_{cell_ i}d\mathbf{r}_{i}Z^{i}_{\Lambda, \gamma}(\mathbf{r}_{i},z)\frac{\partial}{\partial r_{\mu}}Z^{i}_{\Lambda^{\prime},\gamma}(\mathbf{r}_{i},z^{\prime})
\end{equation}
\end{widetext}
where $cell_i$ defines the region surrounding the site $i$.

We now need to consider how to carry out the averaging over configurations implicit in 
Eq.(\ref{tilde sigma}). Butler~\cite{butler4} showed in detail how to use the CPA to
accomplish this. The single site nature of this effective medium theory means, however, that
the potentials on the different lattices could only be treated as statistically independent.
We will show how to carry out the averaging using the NLCPA whereby short-ranged correlations
can be naturally included. To this end in the next section we summarise briefly the key
aspects we need from the KKR-NLCPA  together with its incorporation into electronic
density functional theory. Full details can be found in references \onlinecite{derwyn,derwyn2,derwyn dft,r-nlcpa}.

\section{The KKR-NLCPA and Electronic Density Functional Theory}

In general the scattering path operator (SPO) between two sites $i$ and $j$ for an electron, moving 
through an effective medium so that it mimicks the average motion in a disordered system, is given by
\begin{equation}
\label{SPO-eff}
\hat{\tau}^{ij}= \hat{t} \, \delta_{ij} + \sum_{k \neq i}\hat{t}\, (G( \mathbf{R}_{i}- \mathbf{R}_{k}) 
+\delta \hat{G}( \mathbf{R}_{i}- \mathbf{R}_{k})) \hat{\tau}^{kj}.
\end{equation}
Here, all quantities are matrices in angular momentum space and the indices $i$,$j$ run over all sites
in the lattice. The $G( \mathbf{R}_{i}- \mathbf{R}_{k})$'s are structure constants. The effective medium is specified by single site t-matrices $\hat{t}$ and effective
structure constant corrections $\delta \hat{G}( \mathbf{R}_{i}- \mathbf{R}_{j})$. The effective
medium {\it must} be translationally invariant so that $\hat{\tau}^{ij}$ is given in terms of a Brillouin
zone integral.
\begin{equation}
\hat{\tau}^{ij}= \frac{1}{\Omega_{BZ}} \int_{\Omega_{BZ}} d \mathbf{k} [\hat{t}^{-1} -G( \mathbf{k})
- \delta \hat{G}( \mathbf{k}) ]^{-1} e^{ i \mathbf{k} \cdot ( \mathbf{R}_{i}- \mathbf{R}_{j})}.
\end{equation}
In order to establish a tractable procedure for determining the effective medium the NLCPA draws
its chief idea from the dynamical cluster approximation (DCA) for interacting electron 
systems~\cite{hettler,jarrell and krishnamurthy}. This is a coarse graining of $\delta \hat{G}$
consistently in real and reciprocal space and a mapping to a self-consistently embedded impurity
cluster problem with appropriate boundary conditions imposed~\cite{gary}. The full translational symmetry
of the underlying lattice is preserved.  The size of the cluster sets the range of correlations that can be 
included. The lattice is divided into `tiles' centred on a superlattice vectors $\mathbf{R}_{C}$ and
each contains $N_c$ sites at positions $\mathbf{R}_{C} + \mathbf{R}_{I},I=1,\cdots N_c$. The Brillouin zone
is also broken into $N_c$ tiles, of volume $\Omega_t= \Omega_{BZ}/N_c$, centred on the cluster momenta $ \mathbf{K}_n ,n=1,\cdots N_c$ and the 
$\mathbf{R}_{I}$'s and $\mathbf{K}_n$ satisfy the following equation:
\begin{equation}
\frac{1}{N_c} \sum_{\mathbf{K}_n} e^{i \mathbf{K}_n \cdot (\mathbf{R}_{I}- \mathbf{R}_{J})} = \delta_{IJ}.
\end{equation}
$\delta \hat{G}( \mathbf{k})$ is coarse-grained so that it has the average value 
$\delta \hat{G}( \mathbf{K}_n )$ in a tile centred on $ \mathbf{K}_n$ and in real space, 
$\delta \hat{G}( \mathbf{R}_{I}- \mathbf{R}_{J}) = (1/N_c) \sum_{\mathbf{K}_n} 
\delta \hat{G}( \mathbf{K}_n ) e^{i \mathbf{K}_n \cdot (\mathbf{R}_{I}- \mathbf{R}_{J})}$ with
$\delta \hat{G}( \mathbf{K}_n )= \sum_{J \neq I} \delta \hat{G}( \mathbf{R}_{I}- \mathbf{R}_{J}) 
e^{-i \mathbf{K}_n \cdot (\mathbf{R}_{I}- \mathbf{R}_{J})}$.

The SPO is coarse-grained,
\begin{equation}
\hat{\tau} (\mathbf{K}_n )= \frac{N_c}{\Omega_{BZ}} \int_{\Omega_t} d \tilde{\mathbf{k}} 
 [\hat{t}^{-1} -G( \tilde{\mathbf{k}}+ \mathbf{K}_n)- \delta \hat{G}( \mathbf{K}_n) ]^{-1}
\end{equation}
appropriate to the reciprocal space tile of volume $\Omega_t$ and in real space for
multiple scattering starting and ending on cluster sites $I$ and $J$ respectively.
\begin{equation}
 \hat{\tau}^{IJ} = \frac{1}{\Omega_{BZ}} \sum_{\mathbf{K}_n} ( \int_{\Omega_t} d \tilde{\mathbf{k}}
[\hat{t}^{-1} -G( \tilde{\mathbf{k}}+ \mathbf{K}_n)- \delta \hat{G}( \mathbf{K}_n) ]^{-1} )
e^{i \mathbf{K}_n \cdot (\mathbf{R}_{I}- \mathbf{R}_{J})}.
\end{equation}
Note how $e^{i \tilde{\mathbf{k}} \cdot (\mathbf{R}_{I}- \mathbf{R}_{J})}$ is taken to be $\approx 1$ as the coarse graining
is applied~\cite{hettler}.  The final step is to find the SPO for an impurity cluster, $\tau^{IJ}_{\gamma}$  describing a particular configuration $\gamma_C$ of atoms which is embedded into the NLCPA medium. By demanding that the average is equal to the SPO of the NLCPA enables the effective t-matrix and structure constant corrections to be determined, i.e.,
 \begin{equation}
\label{NLCPA-ansatz}
\sum_{\gamma_C} P_{\gamma_C} \tau^{IJ}_{\gamma_C} = \hat{\tau}^{IJ}.
\end{equation}
Short-range order, SRO, can be included by choosing the probabilities $P_{\gamma_C}$ appropriately as demonstrated in, for example, refs.~\onlinecite{derwyn2, derwyn dft}.

In ref.~\onlinecite{derwyn dft}, in a generalisation of the work of Johnson et al.~\cite{johnson, johnson-totalE}, it is described how to specify a configurationally averaged electronic Grand potential, $\bar{\Omega}$ in terms of KKR-NLCPA quantities and charge densities $\rho_{\gamma_C} (\mathbf{r}_{I})$
and one-electron potentials, $v_{\gamma_C} (\mathbf{r}_{I})$ different for each cluster
configuration. The functional minimisation of $\bar{\Omega}$ with respect to the charge densities, $\rho_{\gamma_C} (\mathbf{r}_{I})$'s,
determines the total energy of the system and requires the $\rho_{\gamma_C} (\mathbf{r}_{I})$'s and $v_{\gamma_C} (\mathbf{r}_{I})$ 's to be found self-consistently. Rowlands et al~\cite{derwyn dft} applied this DFT to investigate how the total energy, charge densities and densities of states are affected by SRO. Tulip et al.~\cite{spectral} showed how further 
information can be found about the effects of SRO on the electronic structure by formulating and calculating the Bloch spectral function at the cluster momenta and averaged over tiles whilst Batt and Rowlands~\cite{gary}
explained how the spectral function at any point in the Brillouin zone can be found. In the following we build on these developments and describe the theory for a two-particle correlation function of disordered 
system with SRO. The particular example is to the d.c. conductivity.

\section{Analytic configuration averaging of the conductivity using the KKR-NLCPA}

From the above it is clear that the KKR-NLCPA should enable an analytical configurational average of 
the conductivity to be carried out. To do so, some care needs to be exercised.  There are two distinct
cases that we must consider in Eq.(\ref{tilde sigma}): (i) where the
two sites under consideration, $i$ and $j$, lie within the same NLCPA
cluster, and (ii) when they lie in two different clusters, in which case
the occupancies of the sites will be statistically independent, as the
two distinct clusters will be statistically independent. This is a
natural generalisation of Butler's work \cite{butler4}, where he
distinguishes between the two cases of $i=j$ and $i\ne j$. 

Hereon we use lower case letters to denote general sites in the lattice,$i,j,\cdots$; upper case $C$ denotes tiles containing the clusters; and upper case letters, with the exception of $C$, denote sites within clusters. So for a site at position $ \mathbf{R}_i$ we use $\mathbf{R}_i = \mathbf{R}_C + \mathbf{R}_I$ and $I,J, \cdots$ label sites within tile $C$, $I^{\prime},J^{\prime}, \cdots$ sites within tile $C^{\prime}$ etc.

We accordingly write
\begin{equation}
\tilde{\sigma}_{\mu\nu}(z_{1},z_{2})=\tilde{\sigma}^{0}_{\mu\nu}(z_{1},z_{2})+\tilde{\sigma}^{1}_{\mu\nu}(z_{1},z_{2})
\end{equation}

\begin{widetext}
\begin{equation}
\tilde{\sigma}^{0}_{\mu\nu}(z_{1},z_{2})=-\frac{4m^{2}}{\pi\hbar^{3}\Omega}\sum_{J \in C}
\sum_{\Lambda_{1},\Lambda_{2},\Lambda_{3},\Lambda_{4}}<J^{I}_{\Lambda_{1}\Lambda_{2};\mu}(z_{2},z_{1})
\tau^{IJ}_{\Lambda_{2}\Lambda_{3}}(z_{1})J^{J}_{\Lambda_{3}\Lambda_{4};\nu}(z_{1},z_{2})
\tau^{JI}_{\Lambda_{4}\Lambda_{1}}(z_{2})>
\end{equation}
and 
\begin{equation}
\tilde{\sigma}^{1}_{\mu\nu}(z_{1},z_{2})=-\frac{4m^{2}}{\pi\hbar^{3}\Omega} \sum_{C^{\prime} \neq C} \sum_{J^{\prime} \in C^{\prime}}\sum_{\Lambda_{1},\Lambda_{2},\Lambda_{3},\Lambda_{4}}
<J^{I}_{\Lambda_{1}\Lambda_{2};\mu}(z_{2},z_{1})\tau^{I,C^{\prime}+J^{\prime}}_{\Lambda_{2}\Lambda_{3}}(z_{1})
J^{J^{\prime}}_{\Lambda_{3}\Lambda_{4};\nu}(z_{1},z_{2}) \tau^{C^{\prime}+J^{\prime},I}_{\Lambda_{4}\Lambda_{1}}(z_{2})>
\end{equation}
\end{widetext}
where $\tilde{\sigma}^{0}$ includes sites $J$ within the same NLCPA
cluster (denoted $C$) as our reference site $I$, and $\tilde{\sigma}^{1}$ includes
all sites lying outside this cluster. Note that in writing these
equations, we have utilised the translational invariance of the
averaged system to remove the second sum appearing in Eq.(\ref{tilde sigma}).

We now introduce response functions such that we can write
\begin{widetext}
\begin{equation}
\label{average}
\tilde{\sigma}^{0}_{\mu\nu}(z_{1},z_{2})=-\frac{4m^{2}}{\pi\hbar^{3}\Omega} \sum_{\gamma_C}P_{\gamma_C} \sum_{\Lambda_{1},\Lambda_{2}}
J^{I,\gamma_C}_{\Lambda_{1}\Lambda_{2};\mu}(z_{2},z_{1})
K^{I;C,\gamma_C}_{\Lambda_{2},\Lambda_{1}; \nu}(z_{1},z_{2})
\end{equation}

\begin{equation}
\label{original K response}
K^{I; C, \gamma_C}_{\Lambda_{2},\Lambda_{1}; \nu}(z_{1},z_{2})=\sum_{J\in C}
\sum_{\Lambda_{3},\Lambda_{4}}<\tau^{IJ}_{\Lambda_{2}\Lambda_{3}}(z_{1})
J^{J,\gamma_C,\nu}_{\Lambda_{3}\Lambda_{4}}(z_{1},z_{2})
\tau^{JI}_{\Lambda_{4}\Lambda_{1}}(z_{2})>_{C,\gamma_C}.
\end{equation}
\end{widetext}
$K^{I; C, \gamma_C}_{\Lambda_{2},\Lambda_{1}; \nu}(z_{1},z_{2})$ involves an average over all configurations
except the configuration is fixed in cluster $C$ to be $\gamma_C$. The single site quantity $J^{I,\gamma_C,}_{\Lambda_{1}\Lambda_{2};\mu}(z_{2},z_{1})$ is set up depending on what kind of element occupies site $I$ and the one-electron potential $v_{\gamma_C} (\mathbf{r}_{I})$ that is dependent on the configuration $\gamma_C$.

Similarly, for the inter-cluster contributions to the conductivity, we
can introduce the following
\begin{widetext}
\begin{equation}
\label{inter-cluster}
\tilde{\sigma}^{1}_{\mu\nu}(z_{1},z_{2})  =  -\frac{4m^{2}}{\pi\hbar^{3}\Omega} 
\sum_{C^{\prime} \neq C}  \sum_{\gamma_C} P_{\gamma_C} \sum_{\gamma_{C^{\prime}}} P_{\gamma_{C^{\prime}}} \sum_{\Lambda_{1},\Lambda_{2}}
J^{I,\gamma_C}_{\Lambda_{1}\Lambda_{2};\mu}(z_{2},z_{1})
L^{I,C,\gamma_C;C^{\prime},\gamma_{C^{\prime}}}_{\Lambda_{2},\Lambda_{1}; \nu}(z_{1},z_{2})
\end{equation}
and
\begin{equation}
\label{l average}
L^{I,C,\gamma_C; C^{\prime},\gamma_{C^{\prime}}}_{\Lambda_{2},\Lambda_{1}; \nu}(z_{1},z_{2})=
\sum_{J^{\prime}\in C^{\prime}}\sum_{\Lambda_{3},\Lambda_{4}}<\tau^{I,C^{\prime}+J^{\prime}}_{\Lambda_{2}\Lambda_{3}}(z_{1})
J^{J^{\prime},\gamma_{C^{\prime}}}_{\Lambda_{3}\Lambda_{4};\nu}(z_{1},z_{2})
\tau^{C^{\prime}+J^{\prime},I}_{\Lambda_{4}\Lambda_{1}}(z_{2})>_{C,\gamma_C;C^{\prime},\gamma_{C^{\prime}}} 
\end{equation}
\end{widetext}  
where the notation is similar to before but the average now fixes cluster $C$ to be occupied by configuration $\gamma_C$  and cluster $C^{\prime}$ to be loaded with configuration $\gamma_{C^{\prime}}$. 

In order to evaluate the ensemble averages contained in Eqs.(\ref{average})-(\ref{l average}) it is helpful to express the SPO, $\tau$, for a particular
configuration as the SPO in the NLCPA medium plus corrections. Again this is the direct generalisation of Butler's approach~\cite{butler4}. We can write
\begin{equation}
\label{scattering equation}
\tau^{ij}=\hat{\tau}^{ij}+\sum_{k,l}\hat{\tau}^{ik}T^{kl}\hat{\tau}^{lj}
\end{equation}
where the effective medium path SPO is denoted
$\hat{\tau}$  as before and $T$ is the total scattering
matrix relevant to the specific configuration. The double summation is taken over all lattice sites.
 $\tau^{ij}$ satisfies the following equation 
\begin{equation}
\sum_{k}(t^{-1}_{k}\delta_{ik}-G( \mathbf{R}_{i}- \mathbf{R}_{k}))\tau^{kj}=\delta_{ij}
\end{equation}
where all quantities are matrices in angular momentum space.  We 
consider fluctuations about the NLCPA medium to obtain
\begin{equation}
\sum_{k}\bigg( (t^{-1}_{k}-\hat{t}^{-1})\delta_{ik} + \delta \hat{
  G}( \mathbf{R}_{i}- \mathbf{R}_{k})+\hat{t}^{-1}-\delta \hat{
  G}( \mathbf{R}_{i}- \mathbf{R}_{k})-G( \mathbf{R}_{i}- \mathbf{R}_{k})\bigg)\tau^{kj}=\delta_{ij}
\end{equation}
which can be re-arranged to yield
\begin{equation}
\label{tauij}
\tau^{ij}=\hat{\tau}^{ij}-\sum_{k,l}\hat{\tau}^{ik} \Delta m^{kl}\tau^{lj}.
\end{equation}
 with 
\begin{equation}
\label{Delta m}
\Delta m^{kl}=(t^{-1}_{k}-\hat{t}^{-1}) \delta_{kl} - \delta \hat{G}( \mathbf{R}_{k}- \mathbf{R}_{l})
\end{equation}
and have used the fact that the effective medium SPO may be written as the inverse of the matrix with
elements $(\hat{t}^{-1} \, \delta_{ij} -\delta \hat{G}( \mathbf{R}_{i}- \mathbf{R}_{j}) -G ( \mathbf{R}_{i}- \mathbf{R}_{j}))$.

We can thus write down
\begin{equation}
\sum_{l}T^{kl}\hat{\tau}^{lj}=\sum_{l}\Delta m^{kl} \tau^{lj}.
\end{equation}
If we now
substitute for $\tau^{lj}$ using Eq.( \ref{scattering equation}), label the sites according to
clusters, $C$ and sites within those clusters (uppercase letters), we obtain
\begin{equation}
T^{C+K,C^{\prime}+L^{\prime}}=x^{KL} \delta_{C,C^{\prime}}+\sum_{C^{\prime \prime}\neq C} \sum_{M N^{\prime \prime}} x^{KM}\hat{\tau}^{C+M,C^{\prime \prime}+N^{\prime \prime}}T^{C^{\prime \prime}+N^{\prime \prime},C^{\prime}+L^{\prime}}
\end{equation}
where we have introduced the matrix $x$ associated with a single cluster of sites, given by
\begin{equation}
\label{x}
x^{IJ} =  \sum_{K} \bigg( 1 + \Delta m \, \hat{\tau}\bigg)^{-1}_{IK} \Delta m^{KJ}.
\end{equation}

Our results here
are a direct cluster generalisation of Butler's \cite{butler4}. Note also that these results
are consistent with Hwang \emph{et al}'s cluster CPA conductivity formalism~\cite{hwang} (although
that is phrased in terms of $t$-matrices, rather than the $x$-matrix
that we use in this work). Further, the special case of a single-site
cluster recovers the more familiar CPA results. 

The NLCPA amounts to writing 
\begin{widetext}
\begin{equation} 
<T^{C+K,C^{\prime}+L^{\prime}}>_{NLCPA}=<x^{KL}> \delta_{C,C^{\prime}} +\sum_{C^{\prime \prime}\neq C}\sum_{M N^{\prime \prime}}<x^{KM}>\hat{\tau}^{C+M, C^{\prime \prime} +N^{\prime \prime}}<T^{C^{\prime \prime}+N^{\prime \prime}, C^{\prime}+L^{\prime}}>_{NLCPA}
\end{equation}
\end{widetext}
and if we choose that $<x^{IJ}>=0$, which is another way of expressing the NLCPA ansatz (Eq. (\ref{NLCPA-ansatz})) then $<T>_{NLCPA}=0$, and we
obtain $<\tau>_{NLCPA}=\hat{\tau}$, (Eq. (\ref{NLCPA-ansatz})). Of course, in writing this, we
have made the approximation that 
\begin{equation}
<x^{KM}\hat{\tau}^{C+M, C^{\prime \prime}+ N^{\prime \prime}}T^{C^{\prime \prime}+N^{\prime \prime}, C^{\prime} +L^{\prime}}>\approx <x^{KM}>\hat{\tau}^{C+M,C^{\prime\prime}+N^{\prime \prime}}<T^{C^{\prime\prime}+N^{\prime \prime},C^{\prime}+L^{\prime}}>
\end{equation}
which is analogous to the usual CPA-type averaging approximation. 

Using Eq. (\ref{scattering equation}) and closely following a cluster generalisation of Butler's
analogous derivation, which refers to fluctuations about the single site CPA medium, we find (suppressing angular 
momentum labelling)
\begin{equation}
\label{K-Ds}
K^{I; C,\gamma_C}_{\nu}=
\sum_{M,N}D^{\gamma_C}_{IM}\tilde{K}_{MN;\nu}^{C,\gamma_C}D^{\dagger \gamma_C}_{NI}
\end{equation}
with
\begin{equation}
\label{compact intra cluster}
\tilde{K}_{IJ;\nu}^{C,\gamma_C}=\sum_{KL} \hat{\tau}^{IK}\tilde{J}_{KL;\nu}^{\gamma_C}\hat{\tau}^{LJ}
+\sum_{C^{\prime \prime} \neq C}\sum_{K^{\prime \prime},L^{\prime \prime}}\hat{\tau}^{I,C^{\prime \prime} +K^{\prime \prime}}\Gamma^{C,\gamma_C}_{C^{\prime \prime};K^{\prime \prime}L^{\prime \prime};\nu} \hat{\tau}^{C^{\prime \prime} +L^{\prime \prime},J}
\end{equation}
and
\begin{equation}
\label{L-Ds}
L^{I,C,\gamma_C; C^{\prime} \gamma_{C^{\prime}}}_{\nu}=
\sum_{M,N}D^{\gamma_C}_{IM}\tilde{L}_{MN;\nu}^{C,\gamma_C; C^{\prime},\gamma_{C^{\prime}}} D^{\dagger \gamma_C}_{NI}
\end{equation}
where
\begin{eqnarray}
\label{L tilde}
\tilde{L}_{MN;\nu}^{C,\gamma_C; C^{\prime},\gamma_{C^{\prime}}} & = & \sum_{K^{\prime},L^{\prime}}\hat{\tau}^{M,C^{\prime}+K^{\prime}}
\tilde{J}^{\gamma_{C^{\prime}}}_{K^{\prime}L^{\prime};\nu}\hat{\tau}^{C^{\prime}+L^{\prime}}\nonumber
\\
& & +\sum_{C^{\prime \prime}\ne(C,C^{\prime})}\sum_{K^{\prime \prime},L^{\prime \prime}}(\hat{\tau}^{M, C^{\prime \prime}+K^{\prime \prime}}\Gamma^{C,\gamma_C;C^{\prime},\gamma_{C^{\prime}}}_{C^{\prime \prime},K^{\prime \prime},L^{\prime \prime};\nu}\hat{\tau}^{C^{\prime \prime}+L^{\prime \prime},N}).
\end{eqnarray}

In Eqs. (\ref{K-Ds}) and (\ref{L-Ds}) we use the NLCPA projector $D$($D^{\dagger}$) which is
$D=(1+\Delta m \hat{\tau})^{-1}$ ($(1+\hat{\tau} \Delta m)$) found in Eqs. (\ref{Delta m}) and (\ref{x}).
The NLCPA ansatz can be re-written in terms of them, i.e. $\sum_{\gamma_C} P_{\gamma_C} D^{\gamma_C} =1$. We have also defined the current quantities
\begin{equation}
\label{tilde J}
\tilde{J}^{\gamma_C}_{KL;\nu}=\sum_{N} D^{\dagger \gamma_C}_{K N}J^{N,\gamma_C}_{\nu} D^{\gamma_C}_{NL}.
\end{equation}
in Eqs.(\ref{compact intra cluster}) and (\ref{L tilde}). Finally
we have introduced vertex functions $\Gamma^{C,\gamma_C}_{C^{\prime \prime},K^{\prime \prime},L^{\prime \prime}; \nu}$,
$\Gamma^{C,\gamma_C;C^{\prime},\gamma_{C^{\prime}}}_{C^{\prime \prime},K^{\prime \prime},L^{\prime \prime};\nu}$
which are the NLCPA analogues of the vertex functions derived by Butler~\cite{butler4}. We now show how to calculate these quantities. 

\section{Vertex Functions and the Conductivity} 

Our two vertex functions are slightly different. $\Gamma^{C,\gamma_C}_{C^{\prime \prime};K^{\prime \prime},L^{\prime \prime};\nu}$, which appears in Eq.(\ref{compact intra cluster}) for the intra-cluster component of
the conductivity, concerns the connection between the configurational occupation in one cluster, $C$, with that of another, $C^{\prime \prime}$, whereas the vertex function $\Gamma^{C,\gamma_C;C^{\prime},\gamma_{C^{\prime}}}_{C^{\prime \prime},K^{\prime \prime},L^{\prime \prime};\nu}$ for the inter-cluster contribution relates the contents of cluster $C^{\prime \prime}$ with that in two others, the reference one, $C$ and another $C^{\prime}$.  To facilitate the derivation of a closed set of equations, we introduce
an approximation, analogous to that in ref. ~\onlinecite{butler4}, and assume that the dependence of the latter on the contents of cluster $C$ may be neglected. Thus
\begin{equation}
\label{new vertex function}
\Gamma^{C,\gamma_C;C^{\prime},\gamma_{C^{\prime}}}_{C^{\prime \prime},K^{\prime \prime},L^{\prime \prime};\nu}=\Gamma^{C^{\prime},\gamma_{C^{\prime}}}_{C^{\prime \prime},K^{\prime \prime},L^{\prime \prime};\nu}.
\end{equation}
leading to $\tilde{L}_{MN;\nu}^{C,\gamma_C; C^{\prime},\gamma_{C^{\prime}}}=\tilde{L}_{MN;\nu}^{ C^{\prime},\gamma_{C^{\prime}}}$ in Eq.(\ref{L tilde}). Using Eq. (\ref{scattering equation}), the NLCPA condition, Eq.(\ref{NLCPA-ansatz}) (or its equivalent renditions) and tracking the steps in ref.~\cite{butler4} we obtain
\begin{widetext}
\begin{eqnarray}
\Gamma^{C^{\prime},\gamma_{C^{\prime}}}_{C,IJ;\nu} & = &
\sum_{K,L,M^{\prime},N^{\prime}}<x^{IK,\gamma_C}\hat{\tau}^{K,C^{\prime}+M^{\prime}}
\tilde{J}^{\gamma_{C^{\prime}}}_{M^{\prime}N^{\prime};\nu}\hat{\tau}^{C^{\prime}+N^{\prime},L}
x^{LJ,\gamma_C}>_{C^{\prime},\gamma^{C^{\prime}}}\nonumber\\
& & +\sum_{C^{\prime \prime}\neq(C,C^{\prime})}\sum_{K,L,M^{\prime \prime},N^{\prime \prime}}<x^{IK,\gamma_C}\hat{\tau}^{K,C^{\prime \prime} +M^{\prime \prime}}
\Gamma^{C^{\prime},\gamma^{\prime}}_{C^{\prime \prime},M^{\prime \prime}N^{\prime \prime};\nu}\hat{\tau}^{C^{\prime \prime}+N^{\prime \prime},L}x^{LJ,\gamma_C}>_{C^{\prime},\gamma^{C^{\prime}}},
\end{eqnarray}
\end{widetext}
which, if we compare with Eq.(\ref{L tilde}), allows us to write
the vertex function in terms of the response function
\begin{equation}
\Gamma^{C^{\prime},\gamma_{C^{\prime}}}_{C,IJ;\nu}=\sum_{K,L}<x^{IK,\gamma_C}\tilde{L}_{KL;\nu}^{C^{\prime},\gamma_{C^{\prime}}}x^{LJ,\gamma_C} >_{C^{\prime},\gamma_{C^{\prime}}}.
\end{equation}

This yields a closed set of equations for the conductivity. We may now write Eqs. (\ref{average}) and (\ref{inter-cluster}) as
\begin{widetext}
\begin{equation}
\label{cond0}
\tilde{\sigma}^{0}_{\mu\nu}(z_{1},z_{2})=-\frac{4m^{2}}{\pi\hbar^{3}\Omega N_c}\sum_{\gamma_C}P_{\gamma_C}\sum_{\Lambda_{1},\Lambda_{2}}\sum_{I,J}
\tilde{J}^{\gamma_C}_{JI,\Lambda_{2},\Lambda_{1};\mu}(z_{2},z_{1})
\tilde{K}^{C, \gamma_C}_{IJ, \Lambda_{1},\Lambda_{2};\nu}(z_{1},z_{2})
\end{equation}
and 
\begin{eqnarray}
\label{cond1}
\tilde{\sigma}^{1}_{\mu\nu}(z_{1},z_{2}) & = & -\frac{4m^{2}}{\pi\hbar^{3}\Omega N_c}\sum_{I \in C} \sum_{C^{\prime} \neq C} \sum_{J^{\prime} \in
    C^{\prime}} \sum_{\gamma_C,\gamma_{C^{\prime}}} P_{\gamma_C} P_{\gamma_{C^{\prime}}}\times \nonumber \\
 & & \sum_{\Lambda_{1},\Lambda_{2}}\sum_{I,J}
\tilde{J}^{\gamma_C}_{JI,\Lambda_{2},\Lambda_{1};\mu}(z_{2},z_{1})
\tilde{L}^{C^{\prime},\gamma_{C^{\prime}}}_{IJ,\Lambda_{1},\Lambda_{2}; \nu}(z_{1},z_{2}).
\end{eqnarray}

 The response functions that determine the conductivity are given by:
\begin{eqnarray}
\label{K response}
\tilde{K}_{IJ, \Lambda_{1}, \Lambda_{2}; \nu}^{C,\gamma_C} & = & \sum_{K,L} \sum_{\Lambda_{3},\Lambda_{4}}
\hat{\tau}_{\Lambda_{1}\Lambda_{3}}^{IK}
\tilde{J}^{\gamma_C}_{KL,\Lambda_{3}\Lambda_{4};\nu}\hat{\tau}_{\Lambda_{4}\Lambda_{2}}^{LJ} + \\   \sum_{C^{\prime} \neq C}\sum_{K^{\prime},M^{\prime},N^{\prime},L^{\prime}} & \sum_{\Lambda_{3},\Lambda_{4},
\Lambda_{5},\Lambda_{6}}  & \hat{\tau}_{\Lambda_{1}\Lambda_{3}}^{I, C^{\prime}+K^{\prime}}<x_{\Lambda_{3}\Lambda_{4}}^{\gamma_{C^{\prime}},K^{\prime}M^{\prime}}
\tilde{L}_{M^{\prime}N^{\prime},\Lambda_{4},\Lambda_{5} ;\nu}^{C,\gamma_C} x_{\Lambda_{5}\Lambda_{6}}^{\gamma_{C^{\prime}},N^{\prime}L^{\prime}}>_{C,\gamma_C}
\hat{\tau}_{\Lambda_{6}\Lambda_{2}}^{C^{\prime}+L^{\prime},J} \nonumber
\end{eqnarray}

\begin{eqnarray}
\label{transport equation}
\tilde{L}_{IJ, \Lambda_{1}, \Lambda_{2};\nu}^{C^{\prime},\gamma_{C^{\prime}}} & = &  \sum_{K^{\prime},L^{\prime}}\sum_{\Lambda_{3}, \Lambda_{4}}
\hat{\tau}_{\Lambda_{1}\Lambda_{3}}^{I,C^{\prime}+K^{\prime}}
\tilde{J}^{\gamma_{C^{\prime}}}_{K^{\prime}L^{\prime},\Lambda_{3}, \Lambda_{4};\nu} \hat{\tau}_{\Lambda_{4}\Lambda_{2}}^{C^{\prime}+L^{\prime},J}+  \\
\sum_{C^{\prime \prime}\neq(C,C^{\prime})}
\sum_{K^{\prime \prime},L^{\prime \prime},M^{\prime \prime},N^{\prime \prime}} & \sum_{\Lambda_{3},\Lambda_{4},\Lambda_{5},\Lambda_{6}} &
\hat{\tau}_{\Lambda_{1}\Lambda_{3}}^{I,C^{\prime \prime}+K^{\prime \prime}}<x_{\Lambda_{3}\Lambda_{4}}^{\gamma_{C^{\prime \prime}},K^{\prime \prime}M^{\prime \prime}}
\tilde{L}_{M^{\prime \prime},N^{\prime \prime},\Lambda_{4},\Lambda_{5}; \nu}^{C^{\prime},\gamma_{C^{\prime}}}
x_{\Lambda_{5}\Lambda_{6}}^{\gamma_{C^{\prime \prime}},N^{\prime \prime},L^{\prime \prime}}>_{C^{\prime},\gamma_{C^{\prime}}}\hat{\tau}_{\Lambda_{6}\Lambda_{2}}^{C^{\prime \prime}+L^{\prime \prime},J}. \nonumber
\end{eqnarray}
\end{widetext}

\section{Solution of Transport Equation}

The conductivity is determined by solution of the transport equation, Eq.(\ref{transport equation}).
We first see that in the key Eqs. (\ref{cond0}) and (\ref{cond1}) we require the response function
$\tilde{L}_{IJ;\nu}^{C^{\prime},\gamma_{C^{\prime}}}$ averaged over configurations $\gamma_{C^{\prime}}$ that can be assigned to a cluster $C^{\prime}$ in a NLCPA tile located at position $\mathbf{R}_{C^{\prime}}$ and also that we require this to be summed over all clusters $C^{\prime}$. We rewrite Eq. 
(\ref{transport equation}), omitting the quantum numbers, $\Lambda_1$,$\Lambda_2$ etc. for brevity, and denote the averages 
$\sum_{\gamma_{C^{\prime}}} P_{\gamma_{C^{\prime}}}
\tilde{L}_{IJ;\nu}^{C^{\prime},\gamma_{C^{\prime}}}$, $\sum_{\gamma_{C^{\prime}}} P_{\gamma_{C^{\prime}}} \tilde{J}^{\gamma_{C^{\prime}}}_{K^{\prime},L^{\prime};\nu}$ as 
$L_{IJ;\nu}^{C^{\prime}}$ and $J_{K^{\prime},L^{\prime};\nu}$ respectively as well as summing over clusters $C^{\prime}$.
\begin{eqnarray}
\label{Lsum}
\sum_{C^{\prime}}L_{IJ;\nu}^{C^{\prime}} &= &\sum_{C^{\prime}} \sum_{K^{\prime},L^{\prime}} \hat{\tau}^{I,C^{\prime} +K^{\prime}} J_{K^{\prime},L^{\prime};\nu} \hat{\tau}^{C^{\prime} +L^{\prime}, J} -
\sum_{K^{\prime},L^{\prime}} \hat{\tau}^{I,K^{\prime}} J_{K^{\prime},L^{\prime};\nu} \hat{\tau}^{L^{\prime}, J} +
\nonumber \\ 
& & \sum_{C^{\prime}} \sum_{C^{\prime \prime} \neq C^{\prime}} 
\sum_{K^{\prime \prime},L^{\prime \prime},M^{\prime \prime},N^{\prime \prime}}
\hat{\tau}^{I,C^{\prime \prime}+K^{\prime \prime}} w^{K^{\prime \prime},L^{\prime \prime},M^{\prime \prime},N^{\prime \prime}} \hat{\tau}^{C^{\prime \prime}+L^{\prime \prime}, J} L_{C^{\prime \prime},{M^{\prime \prime},N^{\prime \prime}; \nu}}^{C^{\prime}}
\end{eqnarray}
where $w^{K^{\prime \prime},L^{\prime \prime},M^{\prime \prime},N^{\prime \prime}}= < x^{K^{\prime \prime},L^{\prime \prime}} x^{M^{\prime \prime},N^{\prime \prime}}>$ and $L_{IJ;\nu}^C$ is defined as zero.

We now write the SPOs in terms of their lattice Fourier transforms, i.e.,
\begin{equation}
\hat{\tau}^{I,C^{\prime} +J^{\prime}} = \frac{1}{\Omega_{BZ}} \sum_{\mathbf{K}_n} \int_{\Omega_t} d
\tilde{\mathbf{k}}\hat{\tau} (\mathbf{K}_n;\tilde{\mathbf{k}}) e^{i (( \mathbf{K}_n +\tilde{\mathbf{k}}) \cdot
(\mathbf{R}_I - \mathbf{R}_{C^{\prime}} - \mathbf{R}_{J^{\prime}})}
\end{equation}
where 
\begin{equation}
\hat{\tau} (\mathbf{K}_n;\tilde{\mathbf{k}}) = [ \hat{t}^{-1} -G( \tilde{\mathbf{k}}+ \mathbf{K}_n)- \delta \hat{G}( \mathbf{K}_n) ]^{-1}.
\end{equation}
We also write the response functions $L_{IJ;\nu}^{C^{\prime}}$ in terms  of cluster lattice Fourier transforms, i.e.,
\begin{equation}
L_{IJ;\nu}^{C^{\prime}} = \frac{N_c}{\Omega_{BZ}} \int_{\Omega_t} d\tilde{\mathbf{k}} \, L_{IJ;\nu} (\tilde{\mathbf{k}}) e^{i \tilde{\mathbf{k}}\cdot (\mathbf{R}_C -\mathbf{R}_{C^{\prime}})}.
\end{equation}
On carrying out the sums over $C$ and $C^{\prime}$ in Eq. (\ref{Lsum}) we thus obtain
\begin{eqnarray}
L_{IJ;\nu} (0) = \sum_{KL} &[\frac{1}{\Omega_{BZ}} \sum_{\mathbf{K}_n ,\mathbf{K}_{n^{\prime}} } &
\int_{\Omega_t} d\tilde{\mathbf{k}} \hat{\tau} (\mathbf{K}_n;\tilde{\mathbf{k}})
e^{i(\mathbf{K}_n+\tilde{\mathbf{k}}) \cdot (\mathbf{R}_I - \mathbf{R}_K)}
 \hat{\tau} (\mathbf{K}_{n^{\prime}};\tilde{\mathbf{k}})
e^{i(\mathbf{K}_{n^{\prime}}  +\tilde{\mathbf{k}}) \cdot (\mathbf{R}_L - \mathbf{R}_J)} \nonumber \\
& & - \hat{\tau}^{IK} \hat{\tau}^{LJ}] J_{KL;\nu} \nonumber \\
 + \sum_{K,L,M,N} w^{K,L,M,N} & [ \frac{1}{\Omega_{BZ}} \sum_{\mathbf{K}_n ,\mathbf{K}_{n^{\prime}} }&
\int_{\Omega_t} d\tilde{\mathbf{k}} \, \hat{\tau} (\mathbf{K}_n;\tilde{\mathbf{k}})
e^{i(\mathbf{K}_n+\tilde{\mathbf{k}}) \cdot (\mathbf{R}_I - \mathbf{R}_K)} \,
 \hat{\tau} (\mathbf{K}_{n^{\prime}};\tilde{\mathbf{k}})
e^{i(\mathbf{K}_{n^{\prime}}  +\tilde{\mathbf{k}}) \cdot (\mathbf{R}_L - \mathbf{R}_J)} \nonumber \\
 & & - \hat{\tau}^{IK} \hat{\tau}^{LJ}] L_{MN;\nu} (0)
\end{eqnarray}
Since $L_{IJ;\nu} (0)$ is translationally invariant, $L_{IJ,\nu} (0) = L_{I+I_1+C_1,J+I_1+C_1;\nu} (0)$ for
a translation by an arbitrary lattice vector $\mathbf{R}_i= \mathbf{R}_{C_1} +\mathbf{R}_{I_1}$ and we can sum it over the $N_c$
tile lattice vectors, $L_{IJ\nu} (0) = \frac{1}{N_c} \sum_{I_1} L_{I+I_1 ,J+ I_1;\nu}$. Using this manipulation and applying the NLCPA coarse graining again so that $e^{i \tilde{\mathbf{k}} \cdot \mathbf{R}_{I}} \approx 1$ we find
\begin{equation}
L_{IJ;\nu} (0) = \sum_{K,L} \chi^{I,K,L,J} J_{KL;\nu} + \sum_{K,L,M,N} w^{K,L,M,N} \chi^{I,K,L,J} L_{MN;\nu} (0)
\end{equation}
with the $\chi$ involving a convolution integral over the Brillouin zone,
\begin{equation}
\chi^{I,K,L,J} = \frac{1}{\Omega_{BZ}} \sum_{\mathbf{K}_n} \int_{\Omega_t} d\tilde{\mathbf{k}} \hat{\tau} (\mathbf{K}_n;\tilde{\mathbf{k}})
e^{i\mathbf{K}_n \cdot (\mathbf{R}_I - \mathbf{R}_K)}
 \hat{\tau} (\mathbf{K}_n;\tilde{\mathbf{k}})
e^{i\mathbf{K}_n   \cdot (\mathbf{R}_L - \mathbf{R}_J)} - \hat{\tau}^{IK} \hat{\tau}^{LJ}.
\end{equation}
$L_{IJ} (0)$ is extracted by inverting a `super' matrix, $[ 1- w \chi]$ which has dimension $N_c \times N_c \times N_{\Lambda} \times N_{\Lambda}$, ($N_{\Lambda}$ specifying the number of angular momentum quantum numbers).

By comparing Eqs. (\ref{K response}) and (\ref{transport equation}) and using the definition $L^{C}_{IJ; \nu} =0$, the second term on the RHS of Eq. (\ref{K response}) vanishes.  We thus find the inter-cluster contribution to the conductivity to be
\begin{widetext}
\begin{equation}
\tilde{\sigma}^{1}_{\mu\nu}  =  -\frac{4m^{2}}{\pi\hbar^{3}\Omega}\sum_{I,J,K,L}\sum_{\Lambda_{1},\Lambda_{2}, \Lambda_{3},\Lambda_{4}}
J_{LI,\Lambda_{1},\Lambda_{2};\mu} \bigg([1-\chi(\mathbf{0}) w]^{-1}
\chi(\mathbf{0})\bigg)^{IJKL}_{\Lambda_{1},\Lambda_{3},\Lambda_{4},\Lambda_{2}}
J_{JK,\Lambda_{3},\Lambda_{4} ;\nu},
\end{equation}
\end{widetext}
and the intra-cluster component to be
 \begin{equation}
\tilde{\sigma}^{0}_{\mu\nu} =\frac{4m^{2}}{\pi\hbar^{3}\Omega} \sum_{\gamma_C} P_{\gamma_C}
\sum_{I,K,L}\sum_{\Lambda_{1},\Lambda_{2},\Lambda_{3},\Lambda_{4}} 
J^{\gamma_C}_{I,\Lambda_{4},\Lambda_{1};\mu} \hat{\tau}^{IK}_{\Lambda_{1},\Lambda_{2}}
\tilde{J}^{\gamma_C}_{KL,\Lambda_{2},\Lambda_{3} ;\nu} \hat{\tau}^{KI}_{\Lambda_{3},\Lambda_{4}}.
\end{equation}

\section{Comparison with the CPA - alloys with no short-range order.}

We have implemented the formalism outlined above using the Munich
self-consistent, spin polarised, relativistic KKR (SPRKKR) code of Ebert \emph{et al}.~\cite{hubert code}. Throughout we use an angular
momentum cutoff $l_{max} = 3$ which is necessary for studies of transition metal systems
with significant d-electron weight in the electronic structure close to the
Fermi energy. The current matrix elements which occur in the conductivity expression
have odd parity and couple  for example d-states to both p- and f-states. Omission of the effect
of the latter can lead to an underestimate of the conductivity.~\cite{butler4,swihart}.
Although in principle the recently developed SCF-KKR-NLCPA method~\cite{derwyn dft} can provide the appropriate 
self-consistent one-electron charge densities and potentials, $\rho_{\gamma_C} (\mathbf{r}_{I})$'s and $v_{\gamma_C} (\mathbf{r}_{I})$ for our transport calculations, in these first applications we use those generated by the faster, simpler SCF-KKR-CPA method~\cite{johnson, johnson-totalE} in order to explore the new aspects of our theory. 

Owing to the deviation of its resistivity from the common Nordheim ($\rho \sim c(1-c)$) behavior ~\cite{Nordheim} and to earlier extensive studies made of it~\cite{butler2,swihart,banhart3}, 
the $AgPd$ series of solid solutions provides the ideal initial test bed for our new method. We implement the formalism with NLCPA clusters containing 4 sites ($N_C=4$) for this f.c.c. based system. Figure 1 shows the calculations of the resistivities of randomly disordered, substitutional $Ag_{c}Pd_{1-c}$ alloys compared with those calculated using the established CPA formulism of Butler~\cite{butler4}. Both 
implementations are presented with and without the so called `vertex corrections'. We see that there is little difference between the two sets of calculations. Each describes the experimental trends well and the vertex corrections are found to be fairly insignificant for each approach. Both of these aspects have been discussed fully in relation to the CPA calculations in earlier publications (e.g. \cite{Turek, Banhart4}) so will not be discussed further here. Extensive work over 2 decades using the CPA formalism has shown how the resistivities of randomly disordered transition metal alloys can be reliably described. 
The good agreement between our NLCPA results and those from the well established CPA method for these alloys where no short-range order is present is a very satisfactory first test of the new formalism.
\begin{figure}[tbh]
\resizebox{0.9\columnwidth}{!}{\includegraphics*{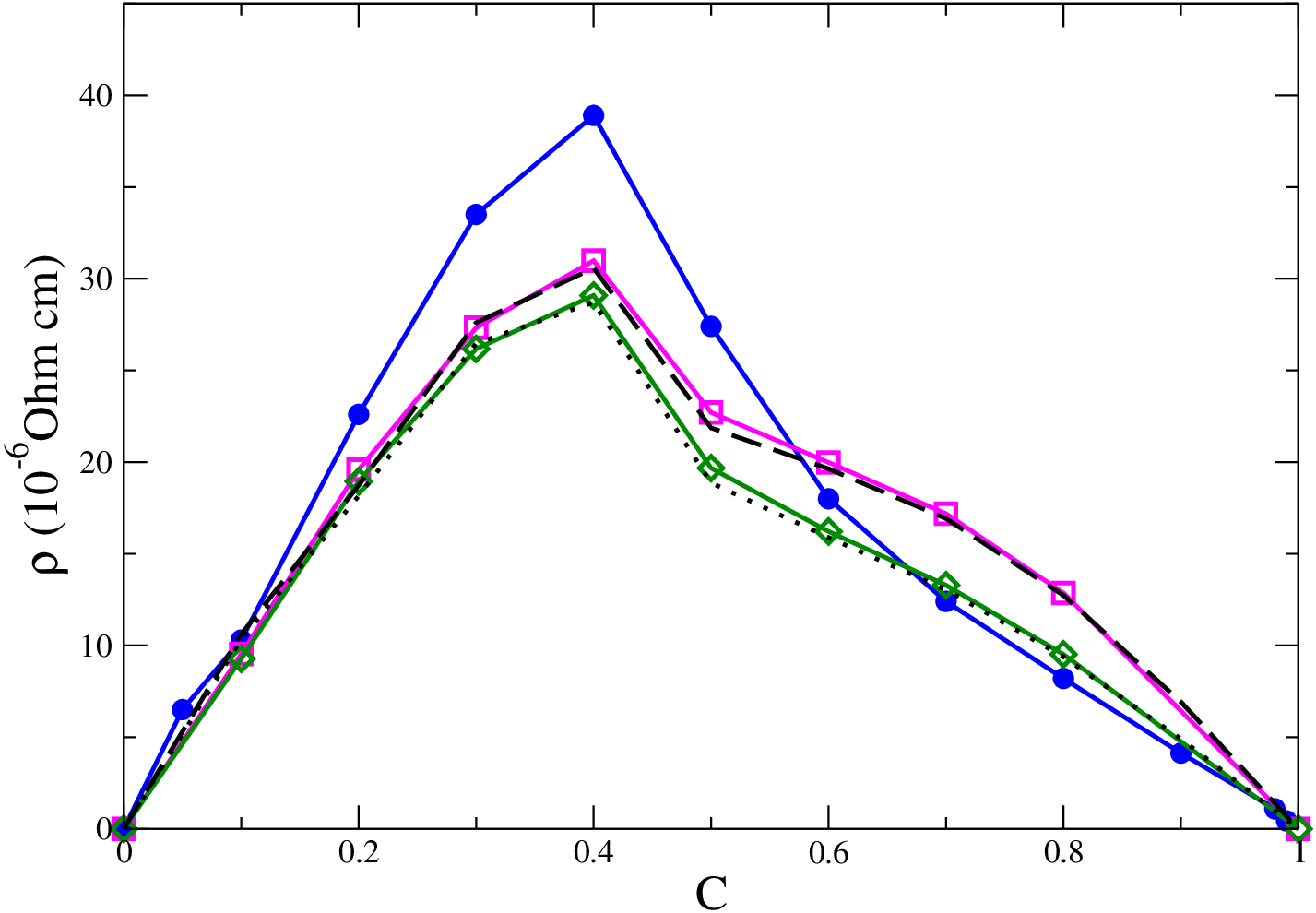}}
\caption{The resistivity of randomly disordered alloys $Ag_{c}Pd_{1-c}$ alloys as a function of concentration $c$. The full lines are the NLCPA results (pink (green) lines with squares(diamonds) show those without (with) vertex corrections). The dashed curves show the CPA results (long dashes (dots) - without (with) vertex corrections). The experimental results of Guenault~\cite{Guenhault} are shown for comparison (full blue lines with filled circles).}
\end{figure}

\section{The effects of SRO on resistivity.}

Many properties of alloys such as resistivity are affected by short-range order.
Indeed resistivity measurements are often used to monitor the changes in SRO which
occur in annealing processes. If an alloy undergoes defect annealing after having
been cold-worked there are significant changes in its physical properties owing
to microstructural changes. For technical applications it is important to know 
what these changes are so that physical properties can be controlled. SRO plays
an important role in this and resistivity measurements are used to follow its 
kinetics.\cite{Migschitz} Our formalism is designed to help the interpretation of such measurements since it can describe the effects of short-range order on transport properties of
alloys. It enables the calculation of the resistivity of a system to be made for a prescribed degree of SRO via the setting of the cluster configurational probabilities $P_{\gamma_{C}}$. Hence it can aid the extraction of SRO attributes from resistivity measurements. 

Our first application is to the b.c.c. based series of disordered $Cu_{c}Zn_{1-c}$ alloys. We implement the NLCPA
resistivity formalism using the smallest clusters and coarsest Brillouin zone tiling, i.e. $N_C =2$. This means correlations only between nearest neighbors can be described.  We
incorporate SRO according to the following 3 prescriptions for the 4 configurational weights, 
$P_1= P(CuCu)$, $P_2= P(CuZn)$, $P_3=P(ZnCu)$ and $P_4= P(ZnZn)$:
\begin{itemize}
\item No SRO, $P(CuCu) = c^2$, $P(CuZn)=P(ZnCu)= c(1-c)$, $P(ZnZn)= (1-c)^2$.
\item Short-range order (minimising number of like nearest neighbors), for $c > 0.5$,
$P(CuCu) = (2c-1)$, $P(CuZn)=P(ZnCu)= (1-c)$, $P(ZnZn)= 0$ and for $c< 0.5$,
$P(CuCu) = 0$, $P(CuZn)=P(ZnCu)= c$, $P(ZnZn)= (1-2c)$.
\item Short-range clustering (maximising number of like nearest neighbors), 
$P(CuCu) = c$, $P(CuZn)=P(ZnCu)= 0$, $P(ZnZn)= (1-c)$.
\end{itemize}
Figure 2 summarises our findings. Vertex corrections are included and are shown to be large for these systems. This concurs with earlier results for the randomly disordered alloys which have low d-electron weight in electronic structure around the Fermi energy. In the absence of any short-range ordering, the results show approximate adherence to the expected Nordheim $c(1-c)$ behavior and the results both with and without vertex corrections are very close to the CPA results and experimental results~\cite{Ho}. Incorporating short-range order
with extent only between nearest neighbors decreases the resistivity as expected for all concentrations.  
The resistivity now follows a rough $c(1-c) + \lambda c^2 (1-c)^2$ dependence so that the greatest reduction occurs at the stoichiometric concentration of $c=0.5$.  Conversely when short-ranged clustering is included the resistivity increases for all
concentrations and the $c(1-c)$ behavior returns. Evidently current is enhanced on both types of atom when they are surrounded by unlike neighbours. These results are in good agreement with experimental 
measurements of the resistivity which show the resistivity to decrease significantly when the alloys are annealed so that short-ranged order is induced.~\cite{Ho}
\begin{figure}[tbh]
\resizebox{0.9\columnwidth}{!}{\includegraphics*{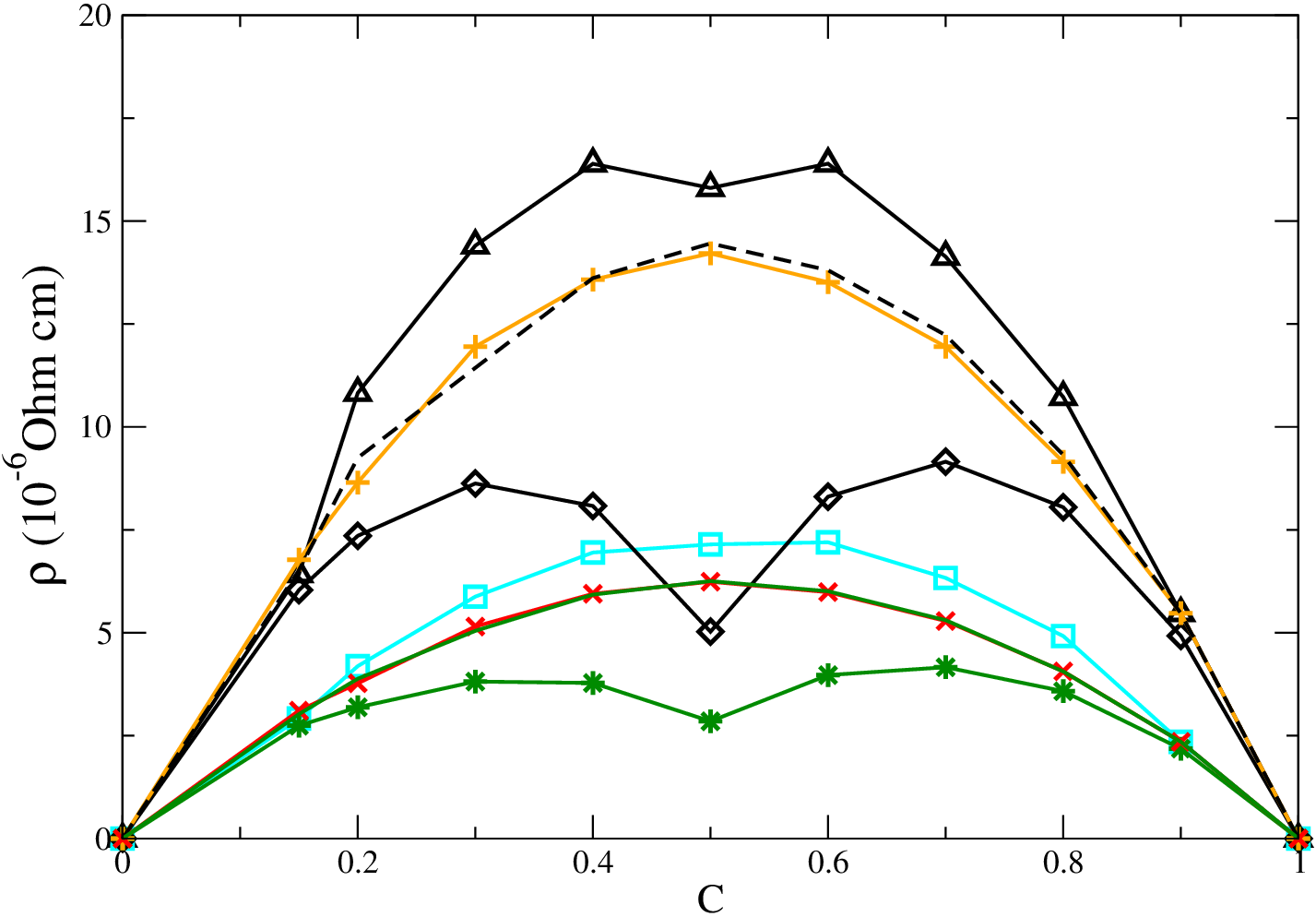}}
\caption{The resistivity of $Cu_{c}Zn_{1-c}$ alloys as a function of $c$, concentration.  The full (red) line with crosses shows the calculations when no SRO is included.(These are nearly indistinguishable from the CPA results (green line)). The  (green) line with asterisks  shows the NLCPA results when SRO is included and the square boxes (light blue line) show results when short-ranged clustering is included. Results are also shown where the vertex corrections have not been included: full (yellow) line with plus signs - NLCPA results for no SRO, (the dashed line shows the CPA results); lines  with diamonds - NLCPA results including SRO;  and lines with triangles - NLCPA results including short-ranged clustering. }
\end{figure}

We have also investigated the effect of short-range order on $Ag_{c}Pd_{1-c}$ alloys by once more 
choosing configurational weights such that the number of like neighbours is minimised for each concentration. Figure 3 contains the results. The effect of SRO is less than that found in $CuZn$. Below $c= 0.3$ the effect is negligible whereas for larger concentrations short-range order depresses the resistivity a little showing how the current is enhanced on a site when surrounded by unlike neighbors. Experimental measurements on $Ag_{c}Pd_{1-c}$ alloys find that annealing has a smaller effect~\cite{Guenhault, Swihart} than in $Cu_{c}Zn_{1-c}$ in line with our calculations. It is also found that cold work causes little change to the resistivity suggesting that additional defects such as dislocations may already be present affecting the measurements. This may indicate the origin of our underestimate of the resistivity shown in Figure 1 and Figure 3. 

\begin{figure}[tbh]
\resizebox{0.9\columnwidth}{!}{\includegraphics*{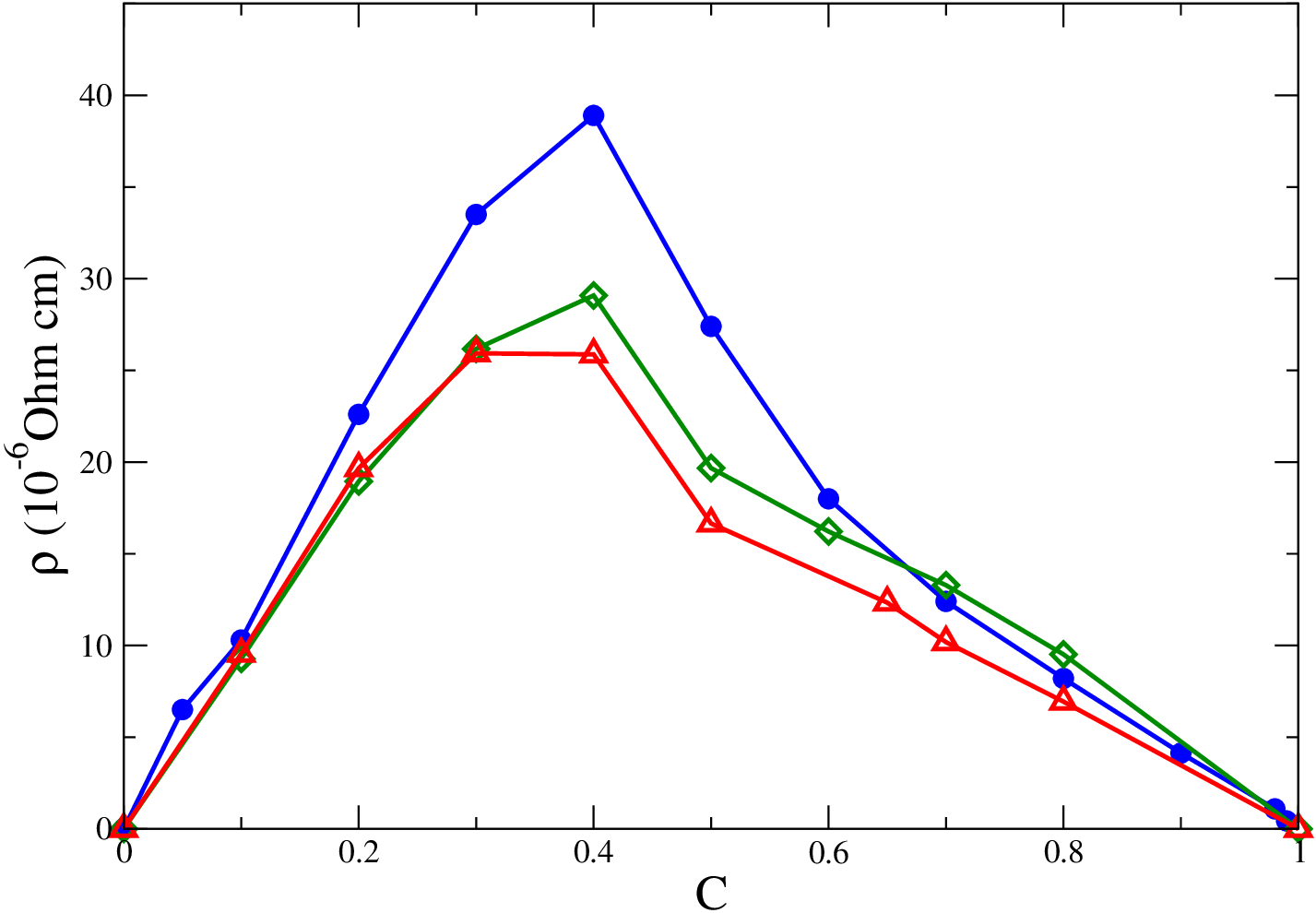}}
\caption{The resistivity of $Ag_{c}Pd_{1-c}$ alloys as a function of $c$, concentration. The full (green) line with diamonds shows the calculations when no SRO is included. The (red) line with triangles  shows those when SRO is included. Vertex corrections are included in both plots. The experimental data~\cite{Guenhault} are also shown (blue line with filled circles).}
\end{figure}

\section{Conclusions}

Short-range ordering and clustering dramatically affect the transport properties of many alloys. Indeed resistivity measurements, which can be made easily and rapidly, provide a good way to monitor microstructural changes that occur in materials processing. In this paper we have described a way
to make quantitative calculations of the resistivity of disordered systems which possess short-range order or clustering. The ab-initio theory starts from the density functional theory for these systems
devised recently by Rowlands et al.~\cite{derwyn dft} using the SCF-KKR-NLCPA electronic structure method. Our first calculations for $Cu_c Zn_{1-c}$ and $Ag_{c}Pd_{1-c}$ show the expected decrease of resistivity when  short-range order is imposed whereas short-ranged clustering produces an increase. For the randomly disordered alloys the results are very similar to those that have been produced from established KKR-CPA resistivity calculations based on Butler et al. work ~\cite{butler4, butler1}.  These first calculations have not included the effects of the short range order on the self-consistent charge densities and potentials that are available from the SCF-KKR-NLCPA method. So far also short-range clustering and ordering effects over only the shortest nearest atomic neighbor range have been included. The computational development work is in progress to remove these current practical limitations. The counterintuitive behavior of the resistivities of the K-state alloys such as $NiCr$,$NiMo$ and $PdW$ will be ideal next systems to study.

\section{Acknowledgements}
The authors are grateful to B. L. Gy\"{o}rffy for very useful discussions.
They also acknowledge financial assistance from the UK
EPSRC and the SSP 1145 "MODERN AND UNIVERSAL
FIRST-PRINCIPLES METHODS FOR MANY-ELECTRON SYSTEMS IN
CHEMISTRY AND PHYSICS" as well as the SFB 689 "Spinph\"anomene in
reduzierten Dimensionen" of the DFG. Computing facilities for this work were provided by the
University of Warwick Centre for Scientific Computing and the Universit\"{a}t M\"{u}nchen.

\end{document}